\documentclass[a4paper,10pt]{iopart}
\usepackage[utf8x]{inputenc}
\usepackage{graphicx}   
\usepackage{amscd}
\usepackage{iopams}
\renewenvironment{pmatrix}{\left(\!\!\begin{array}{cc}}{\end{array}\!\!\right)}
\newcommand{\ri}{{ \rm i }}

\newcommand{\rd}{{ \rm d }}

\newcommand{\be}{\begin{equation}}
\newcommand{\ee}{\end{equation}}

\usepackage{color}
\definecolor{blau}{rgb}{0,0,1}
\definecolor{gruen}{rgb}{0,1,0}
\definecolor{rot}{rgb}{1,0,0}
\definecolor{magenta}{rgb}{1,0,1}

\begin{document}

\title[Tunnelling decay of interacting cold bosons in an optical lattice]{Closed system approach to open systems: Tunnelling decay of interacting cold bosons in an optical lattice}

\author{K. Rapedius} 
\address{FB Physik, TU Kaiserslautern, D-67653 Kaiserslautern, Germany}

\ead{rapedius@physik.uni-kl.de}

\begin{abstract}
A Bose-Hubbard Hamiltonian, modelling cold bosons in an optical lattice, is used to simulate the dynamics of interacting
open quantum systems as subsystems a larger closed system, avoiding complications like the introduction of baths, complex absorbing
potentials or absorbing boundaries. The numerically exact unitary dynamics is compared with effective descriptions of the
subsystems based on non-Hermitian Hamiltonians or Lindblad master equations. The validity of popular models with constant decay rates
is explicitly analyzed for decaying single and double wells. In addition we present a discrete lattice version of the Siegert approximation method
for calculating decay rates.
\end{abstract}

\date{\today}



\maketitle

\section{Introduction}
\label{sec:intro}

The behaviour of interacting quantum particles in open systems is of fundamental interest. Such systems have been studied experimentally
and theoretically in different contexts, including electronic transport in semiconductors and nanostructures \cite{DiVe08} and cavity QED \cite{Scul97}. 

Cold atom experiments \cite{Peth02,Pita03,Bloc08,Bloc12} permit a new way to study interacting quantum particles in open systems and thus a way to test different theoretical
approaches for describing these systems, like effective non-Hermitian Hamiltonians or master equations. Benefits of the cold atom 
approach include the possibility of creating various trap geometries, tunability of the interaction between the particles and the absence 
of defects and impurities. In particular one can realize simple setups where open systems can be studied as smaller subsystems of 
a larger, closed system the dynamics of which is governed by familiar Hermitian Hamiltonians leading to unitary time-evolution. 
In the present article this concept is implemented theoretically for the particular situation of tunnelling decay of cold bosons 
within an optical lattice, modeled by the Bose-Hubbard Hamiltonian. 
\begin{figure}[htb]
\begin{center}
\setlength{\unitlength}{1.1mm}
\begin{picture}(62,32)(4,-4)\thicklines
\multiput(10,10)(10,0){2}{\circle*{4}}
\multiput(30,10)(10,0){3}{\circle{4}}

\multiput(72,10)(10,0){2}{\circle{4}}

\multiput(12,10)(10,0){4}{\line(1,0){6}}

\multiput(64,10)(10,0){2}{\line(1,0){6}}
\multiput(34,12)(10,0){2}{$\Omega$}
\multiput(66,12)(10,0){2}{$\Omega$}

\multiput(54,10)(3,0){3}{\line(1,0){2}}

\multiput(14,12)(10,0){1}{$J$}
\multiput(24,12)(10,0){1}{$\omega$}
\end{picture}
\caption{\label{fig:bild} Bose-Hubbard lattice where two sites (filled circles) forming a double well with internal tunnelling coefficient $J$
 are weakly coupled (coefficient $\omega \ll \Omega$) to a long chain of sites (empty circles) with internal tunnelling coefficient $\Omega$.}
\end{center}
\end{figure}
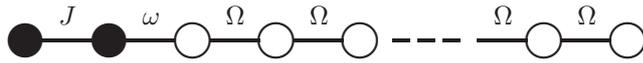

In recent years the decay dynamics of trapped cold bosons was considered in various contexts.
The nonexponential tunnelling decay of Bose-Einstein condensates was analyzed in the Gross-Pitaevskii mean-field approximation for
different trap geometries \cite{Mois04a,Schl06a,Schl06b,08nlLorentz,09ddshell,10nlret}. 
Few boson tunnelling was considered in \cite{Lode09} by means of a numerically exact method yielding deviations from the mean-field
behaviour for small particle numbers.
Furthermore, the decay of bosons in optical lattices has been studied using effective non-Hermitian Hamiltonians \cite{12decoh,08nhbh_s,Hill06},
Lindblad master equations and the Bogoliubov Backreaction approximation \cite{Witt08,Trim11}. In these works penomenological models with localized constant decay rates are used. 
While the latter appear appropriate for describing decay mechanisms like, e.g.~particle loss due to a focused laser beam or electron
beam (the latter was implemented successfully in a recent experiment \cite{Wuer09}),
this does not have to be the case for tunnelling decay considering the nonexponential behaviour found in 
some of the studies mentioned above.

Here we analyze the tunnelling decay of cold bosons in a lattice within the framework of the one-dimensional Bose-Hubbard model 
\be
  {\hat H}= \sum_j \left( \epsilon_j {{\hat a}_j}^\dagger {{\hat a}_j} - \frac{J_j}{2} \left( {{\hat a}_j}^\dagger {\hat a}_{j+1} + {{\hat a}_{j+1}}^\dagger {{\hat a}_j} \right) +  \frac{U_j}{2} {{\hat a}_j}^{\dagger 2} {{\hat a}_j}^2 \right)
\label{BH}
\ee
where ${\hat a}_j$ is the annihilation operator of a particle in the lattice site $j$ and $\epsilon_j$ are on-site single particle energies. 
The interaction parameters $U_j$ depend on the shapes of the local ground states in the respective lattice sites and the local tunnelling rates $J_j$ 
on the overlap of the ground states in the respective adjacent sites \cite{Jaks98}. 

For a high total number of bosons $N$, the operators ${\hat a}_j$ can be replaced by complex numbers $\sqrt{N} c_j$ representing their respective coherent state expectation values
(see e.g.~\cite{Cast00}) which leads to the mean-field Hamiltonian
\be
  H= \sum_j \left( \epsilon_j N |c_j|^2 - \frac{J_j}{2} N \left( c_j^* c_{j+1} + c_{j+1}^* c_j \right) +  \frac{U_j}{2} N (N-1) |c_j|^4 \right) \,.
\label{Ham_GP}
\ee
The dynamics of the on-site amplitudes $c_j$ is then given by $i \dot c_j =\partial H / \partial c_j^*$ which yields
the system of coupled discrete nonlinear Schr\"odinger equations or Gross-Pitaevskii equations
\be
  \ri \dot c_j = \epsilon_j c_j - \frac{J_j}{2} c_{j+1} -\frac{J_{j-1}}{2} c_{j-1} + U_j (N-1) |c_j|^2 c_j 
\label{GP}
\ee
in scaled units with $\hbar=1$ that we will use throughout this article.
In the following the dynamics obtained from numerically exact solutions of equations (\ref{BH}) and (\ref{GP}) is compared with 
the results of effective theoretical models, including non-hermitean Hamiltonians and master equations, describing small subsystems
that are weakly coupled to the rest of the lattice. This is illustrated in figure \ref{fig:bild} for a double well subsystem.

The main objectives of this article can be summed up as follows:
\begin{itemize}
\item An experimentally realizable system, namely cold bosons in a lattice, is used to study non-Hermitian quantum dynamics of interacting particles under clean, 
well-controlled conditions.
\item Effective descriptions of open systems, like e.g.~non-Hermitian Hamiltonians or Lindblad master equations, are compared with
numerically exact calculations within genuinely closed systems without any additional approximations like complex absorbing potentials,
absorbing boundaries or the introduction of particle baths.      
\item The validity of popular phenomenological models with constant decay rates is explicitly tested for a concrete mechanism, namely 
tunnelling  decay within a Bose-Hubbard lattice.
\item The full Bose-Hubbard dynamics is compared with the mean-field approximation.
\item The relative technical simplicity of our approach makes sure that the results are neither obscured nor compromized by mathematical or numerical subtleties.
\item Methodically, the discrete lattice version of the Siegert approximation method is presented as a technically simple 
alternative to decay rate calculations based on Green functions or Fermi's Golden Rule.
\end{itemize}

In section \ref{sec:single} we consider single well tunnelling, i.e.~tunnelling out of one site coupled to a long chain, 
double well tunnelling will be analyzed in section \ref{sec:double}.

\section{Single well tunnelling}
\label{sec:single}
We consider a situation where one site, let us say site $0$, is coupled weakly to a long chain of sites, i.e. we choose the tunnelling
coefficients $J_j=\Omega$, $j>0$ and $J_0=\omega \ll \Omega$. For the sake of simplicity we further choose $\epsilon_0=\epsilon$, $U_0=U$ and $\epsilon_j=0$, $U_j=0$ for $j>0$.
such that there is no interaction within the long chain. 

In the following we will find an approximation to the decay coefficient for tunnelling from the single site into the chain by means of
the Siegert approximation method \cite{Sieg39,08nlLorentz,11siegert} which we adapt for use in a discrete lattice. 

By means of the usual ansatz $c_j(t)=c_j\exp(-\ri \mu t)$ with the chemical potential $\mu$, equation (\ref{GP}) leads to the time-independent nonlinear Schr\"odinger 
equations
\be
   \mu c_0 = -\frac{\omega}{2} c_1 + \epsilon c_0 + U (N-1) |c_0|^2 c_0
\label{c0}
\ee
\be
   \mu c_1 = -\frac{\omega}{2} c_0 -\frac{\Omega}{2} c_2
\label{c1}
\ee
\be
   \mu c_j = -\frac{\Omega}{2} (c_{j-1}+c_{j+1})\,, \qquad j>1 \,.
\label{cj}
\ee
For $j \ge 1$ we make an outgoing wave (Siegert) ansatz $c_j=A \exp(\ri k j)$ in analogy to outgoing plane waves in continuous space. 
Thus equation (\ref{cj}) leads to the dispersion relation $\mu(k)=-\Omega \cos(k)$.
In order to obtain the current in the region $j>1$ we consider $\partial_t|c_j|^2 =\dot c_j^* c_j +c_j^* \dot c_j$ (cf.~e.g. \cite{Gilz11}). 
From (\ref{cj}) we obtain
\be
    \dot{|c_j|^2}= -\ri \frac{\Omega}{2}\left(c_{j+1}^*c_j-c_j^*c_{j+1} + c_{j-1}^*c_j-c_j^*c_{j-1} \right)= -{\cal J}_{j,j+1}-{\cal J}_{j,j-1} 
\ee
where we have identified the currents ${\cal J}_{j,j+1}=\ri \Omega (c_{j+1}^*c_j-c_j^*c_{j+1})/2$ and ${\cal J}_{j,j-1}=\ri \Omega (c_{j-1}^*c_j-c_j^*c_{j-1})/2$
going from site $j$ to sites $j+1$ and $j-1$ respectivelly.
Inserting $c_j=A \exp(\ri k j)$ the ``outgoing'' current becomes ${\cal J}_{j,j+1}=-\Omega |A|^2 \sin(k)=-\Omega |A|^2\sqrt{1-(\mu/\Omega)^2}=:{\cal J}$
where we have used the dispersion relation in the last step. The amplitude $A$ follows from equation (\ref{c1}). Iserting $c_j=A \exp(\ri k j),\, j\ge 1$
and the dispersion relation leads to $A=c_0 \omega/\Omega$. Particle number conservation requires $\dot{|c_0|^2}={\cal J}$. 
If we assume an exponential decay, the decay rate must be given by $\Gamma=-\dot{|c_0|^2}/|c_0|^2={\cal J}/|c_0|^2$ which leads to
\be
  \Gamma=\frac{\omega^2}{\Omega}\sqrt{1-\frac{\mu^2}{\Omega^2}} \, .
\label{Gamma}
\ee
Due to $\omega \ll \Omega$ the term $-\omega c_1/2=-(\omega^2/2 \Omega)c_0 \exp(\ri k)$ in (\ref{c0}) can be neglected so that the chemical 
potential is approximately given by $\mu \approx \epsilon+U(N-1)|c_0|^2$. The occupation of the single site thus satisfies the differential
equation
\be
  \dot{|c_0|^2}=-\frac{\omega^2}{\Omega}\sqrt{1-\frac{(\epsilon+U(N-1)|c_0|^2)^2}{\Omega^2}} |c_0|^2\,.
\label{dotc0Q}
\ee
Alternatively, this result can be obtained by means of Green functions (see, e.g.~\cite{Pesk10}) or Fermi's Golden Rule \cite{Van55}.
Equation (\ref{dotc0Q}) implies that due to the dispersion relation in the lattice there is no decay for 
$|\epsilon+U(N-1)|c_0|^2| \gtrsim |\Omega|$ which can be confirmed numerically.


Now we turn to the full Bose-Hubbard model. We attempt an effective description by means of a master equation for the
probabilities $P_M$ that the single site subsystem is occuppied by $M$ particles, $0 \le M \le N$ \cite{Datt04a}. 
We apply the sequential tunnelling approximation, i.e.~we neglect the possibility of simultaneous decay of two or more 
particles, concentrating on single particle processes only \cite{Datt04a}. 
As in the mean-field limit the transition rates can be obtained using the Siegert approximation method. 
Instead of the mean-field system (\ref{c0})-(\ref{cj}) we consider the Heisenberg equations 
$\ri \dot{{\hat a}_j}=[{\hat a}_j,{\hat H}]$ derived from the Hamiltonian (\ref{BH}).
Since there is no interaction in the chain the current can be obtained in complete analogy with the mean-field case.
The continuity equation still holds so that the tunnelling rate par particle ist still given by (\ref{Gamma}). 
The single site system goes from an $M$-particle state to an $M-1$ particle state when a particle tunnels into the chain. 
At site $0$ the Heisenberg equation reads
\be
  \ri \dot{{\hat a}_0}=-\frac{\omega}{2} {\hat a}_1+ \epsilon_0 {\hat a}_0 + U {\hat a}_0^{\dagger2} {\hat a}_0 \,. 
\ee
The chemical potential for going from state $|M\rangle$ to state $|M-1\rangle$ is approximately determined by its stationary
version
\be
   \mu_M \langle M-1|{\hat a}_0|M\rangle \approx (\epsilon+U(M-1)) \langle M-1|{\hat a}_0|M\rangle 
\ee
where we have neglected the small first term due to $\omega \ll \Omega$ in analogy to the mean-field case. Thus the chemical potential reads 
$\mu_M\approx \epsilon+U(M-1)$. The corresponding tunnelling rates per particle are the given by
\be
   \Gamma_M=\frac{\omega^2}{\Omega}\sqrt{1-\frac{(\epsilon+U(M-1))^2}{\Omega^2}} \, 
\ee
for states with $M$ particles. This leads to the rate equation model
\be
   \dot{P_N}(t)=-N \Gamma_N P_N(t)\,, \quad \dot P_0(t)=\Gamma_1P_1(t) \,,
\ee
\be
   \dot{P_M}(t)=(M+1) \Gamma_{M+1} P_{M+1}(t)-M \Gamma_M P_M(t)\,,\quad 1 \le M \le N-1
\ee
for the relative occupations $P_M$ of the $M$-particle states. The total occupation of the first site is then
given by
\be
   n_0(t)=\sum_{M=0}^N M P_M(t) \, .
\ee
The system can be integrated analytically
\be
   P_N(t)=\exp(-\Gamma_N t)P_N(0),\, \quad P_0(t)=\int_0^t P_1(\tau) \rd \tau \,,
\ee
\be
\fl \qquad   P_{M}(t)=\int_0^t \exp(-M\Gamma_M(t-\tau)) (M+1)\Gamma_{M+1}P_{M+1} \rd \tau\,,\quad 1 \le M \le N-1 \,.
\ee

Now we compare the approximate descriptions obtained above with numerically exact integrations of the full mean-field system and the full Bose-Hubbard 
system for a finite lattice. We start with all particles in the first site. Decay without backreflection can be observed for short times when there is no backreflection at the end of the finite chain. 
The simulation time $T_S$ in a chain of length $L$ is thus determined by $(\Omega/2)T_S\approx L$, $T_S \approx 2L/\Omega$ with the phase 
velocity $\Omega/2$. 
\begin{figure}[htb]
\includegraphics[width=0.49\textwidth] {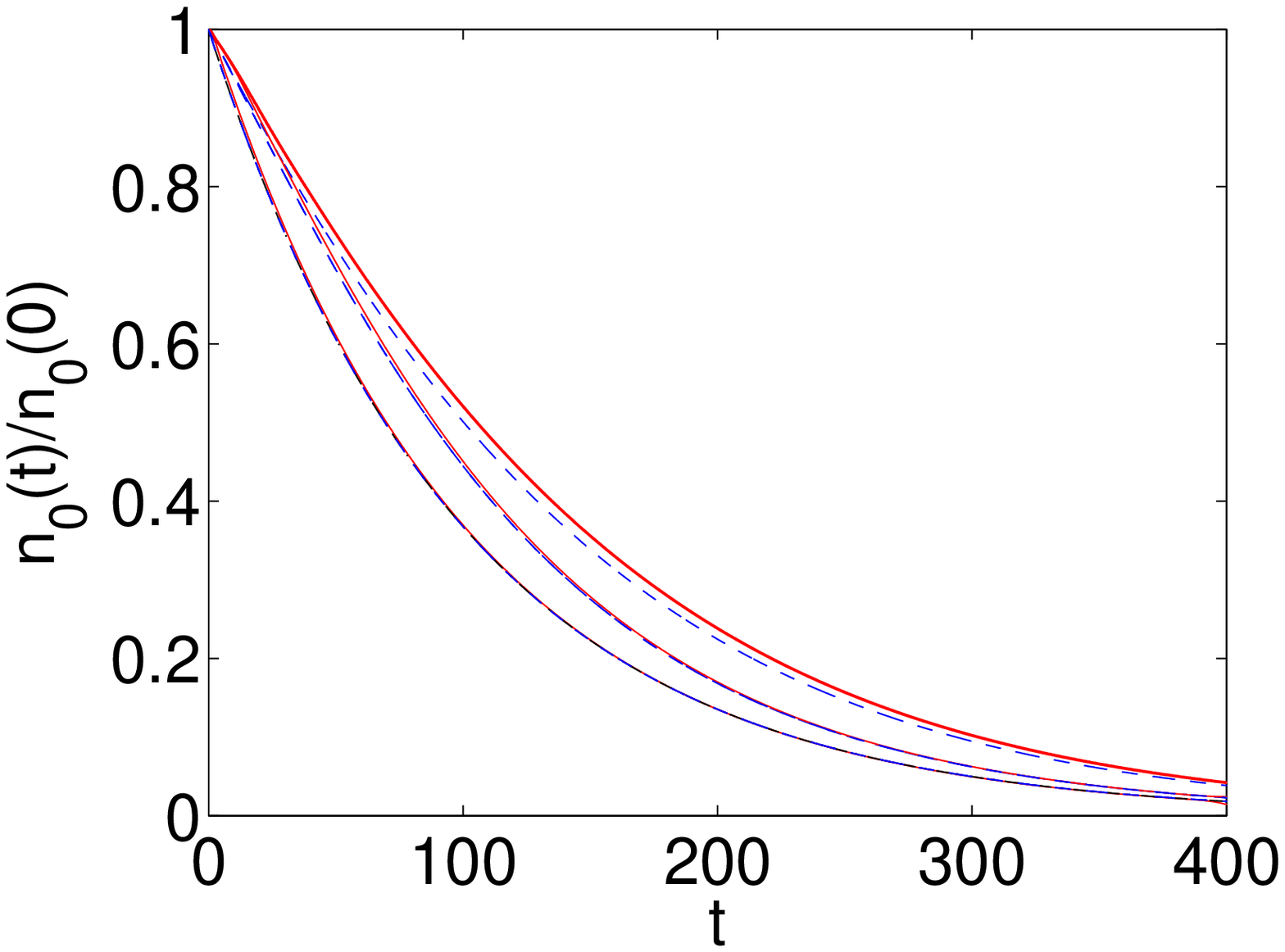}
\includegraphics[width=0.49\textwidth] {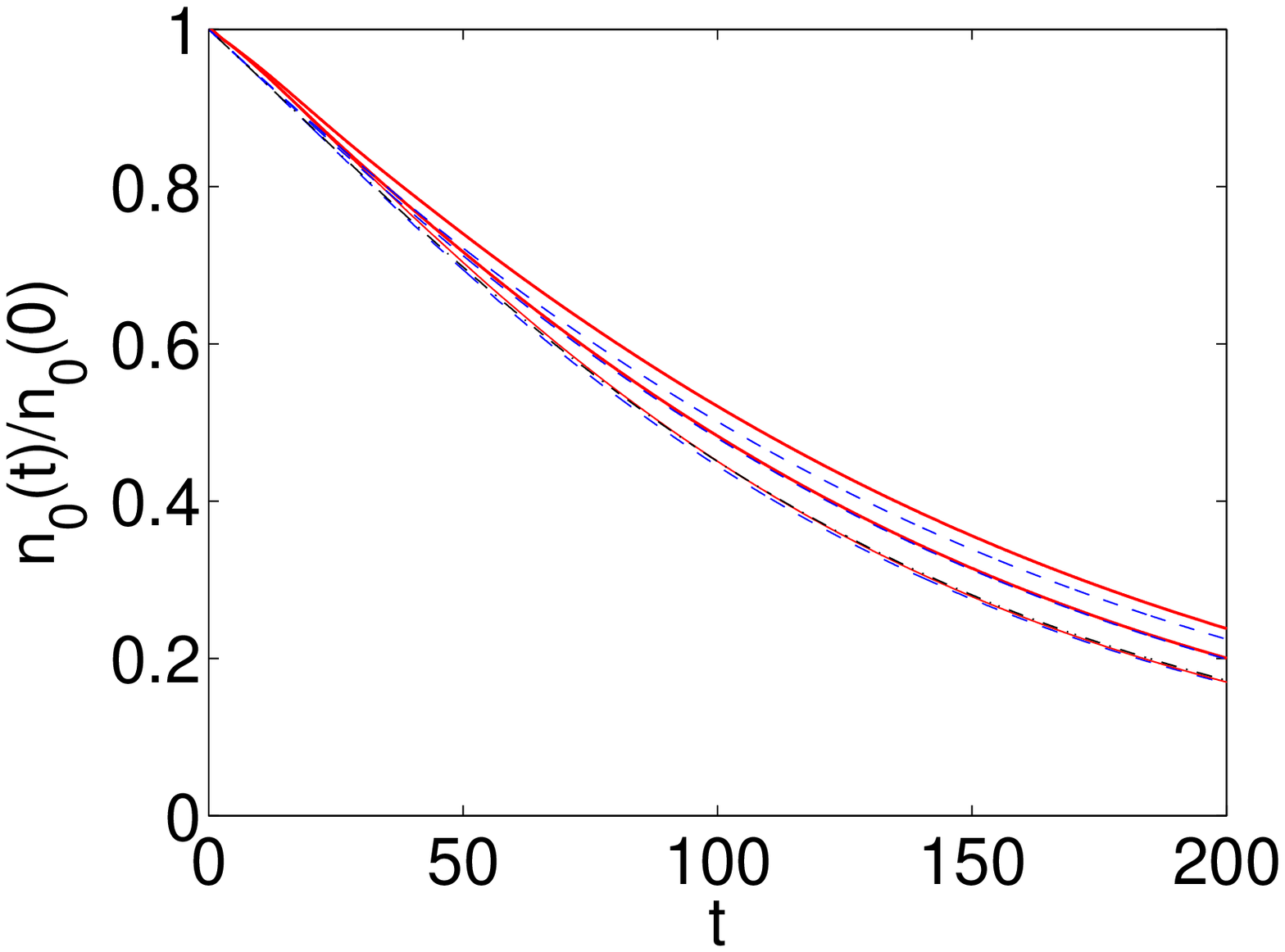}
\caption{\label{fig:single} (Color online) Decay from a single site with parameters $\epsilon=0$, $\omega=0.1$, $\Omega=1$, $\hbar=1$. 
The occupation $n_0(t)/n_0(0)$ of the first lattice site is shown as a function of time. Numerically exact results in a lattice 
 with $200$ (left panels) and $150$ sites (right panels) (solid lines) are compared with the corresponding approximations (dashed lines). Left panel from top to bottom: $N=2$ particles with $U(N-1)=0.8$; mean-field with $U(N-1)=0.8$ and interaction-free system with $U=0$.
Right panel from top to bottom: $N=3$ particles with $U(N-1)=0.8$; $N=2$ particles with $U(N-1)=0.8$; mean-field with $U(N-1)=0.8$; 
dashed-dotted line: Rate equation result for $N=15$ particles with $U(N-1)=0.8$.
}
\end{figure}
The results are shown in figure \ref{fig:single}. In accordance with the results obtained in \cite{Lode09} for single-well tunnelling
in a different (non-lattice) setup, there is a clear difference between few-particle tunnelling and mean-field tunnelling. 
In both cases the numerically exact results (solid lines) are quite well described by the respective approximations discussed above (dashed lines). 
For $N=2$ particles there are some deviations, which might occur due to pair-tunnelling, which has been neglected in the in the rate  
equation model. For both $N=3$ particles and the mean-field there is a much better correspondence between the approximations and the exact 
results, which might be due to the diminished relative importance of pair-tunnelling. For comparison, the right panel also shows the rate equation prediction for $N=15$ particles (dashed dotted line) which 
almost coincides with the mean-field results which indicates that the mean-field limit is approached with relatively few particles.

\section{Double well tunnelling}
\label{sec:double}
Now we consider an open double well, consisting of two sites $0$ and $1$ weakly coupled to a long chain of sites. 
The quasistationary states can again be obtained by means of the Siegert approximation method. The nonlinear Schr\"odinger equations
describing our system in the mean-field limit now read
\be
   \mu c_0 = -\frac{J}{2} c_1 + \epsilon_0 c_0 + U (N-1) |c_0|^2 c_0
\label{c0_2}
\ee
\be
   \mu c_1 = -\frac{\omega}{2} c_2 -\frac{J}{2} c_0 + \epsilon_1 c_1 + U (N-1) |c_1|^2 c_1
\label{c1_2}
\ee
\be
   \mu c_2 = -\frac{\omega}{2} c_1 -\frac{\Omega}{2} c_3
\label{c2_2}
\ee
\be
   \mu c_j = -\frac{\Omega}{2} (c_{j-1}+c_{j+1})\,, \qquad j>2 \,.
\label{cj_2}
\ee
The outgoing wave solution, dispersion relation and current in the chain ($j \ge 2$) are still given by $c_j=A \exp(\ri k j)$, $\mu(k)=-\Omega \cos(k)$ and 
${\cal J}=-\Omega |A|^2\sqrt{1-(\mu/\Omega)^2}$ respectivelly. Analogous to the single site case the amplitude $A$ is obtained from equation
(\ref{c2_2}) as $A=c_1 \omega/\Omega$. The continuity equation $\dot{|c_0|^2}+\dot{|c_0|^2}={\cal J}$ for the first two sites then yields the double well
decay rate
\be
   \Gamma_d= -\frac{{\cal J}}{|c_0|^2+|c_1|^2}=\frac{|c_1|^2}{|c_0|^2+|c_1|^2} \frac{\omega^2}{\Omega}\sqrt{1-\frac{\mu^2}{\Omega^2}} 
\label{Gamma_d}
\ee
where the chemical potential $\mu$ is approximately determined by equations (\ref{c0_2}) and (\ref{c1_2}) the coupling to the chain is neglected, 
i.e.~$\omega$ is set to zero. 
Often, phenomenological models with constant decay rates are used to describe open systems. For our single site model with tunnelling decay
from the previous section this is justified for $\mu \ll \Omega$ when $\Gamma \approx \frac{\omega^2}{\Omega}:=\gamma$. 
Physically this means that the internal dynamics of the subsystem is slow compared to the transport velocity in the chain.
In this limit, one can try to model our decaying double well by simply adding an imaginary constant $-\ri \gamma/2$ to the on-site
energy of the second well which leads to the model
\be
   \ri \dot{c_0} = -\frac{J}{2} c_1 + \epsilon_0 c_0 + U (N-1) |c_0|^2 c_0
\label{c0_2_2}
\ee
\be
   \ri \dot{c_1} =-\frac{J}{2} c_0 + (\epsilon_1-\ri \frac{\gamma}{2} )c_1 + U (N-1) |c_1|^2 c_1
\label{c1_2_2}
\ee
considered in a number of recent works (see, e.~g.~\cite{Witt08,Trim11,08nhbh_s}). It is interesting to compare the predictions of this model for the decay
rates of stationary states with the Siegert approximation result (\ref{Gamma_d}). To this end we treat the term 
$-\ri \gamma c_1 /2$ as a small perturbation. The unperturbed chemical potential and amplitudes $c_0$ and $c_1$ are then
determined by the stationary states of the system (\ref{c0_2_2}) and (\ref{c1_2_2}) with $\gamma$ set to zero, obtained in the usual way
with the ansatz $c_j(t)=\exp(-\ri \mu t)c_j$, $j=0,1$. The first order perturbation theory correction is
straightforwardly obtained as 
\be
 \fl \qquad  \Delta \mu= \frac{1}{|c_0|^2+|c_1|^2}\begin{pmatrix}c_0^* & c_1^*\end{pmatrix}
\begin{pmatrix} 0 & 0 \\0 & -\ri \gamma/2\end{pmatrix}\begin{pmatrix}c_0\\c_1\end{pmatrix}=-\ri \frac{|c_1|^2}{|c_0|^2+|c_1|^2}\frac{\gamma}{2}\,.
\ee
This purely imaginary correction corresponds to a decay rate 
\be
   \Gamma_d'= -2 {\rm Im}(\Delta \mu)=\frac{|c_1|^2}{|c_0|^2+|c_1|^2} \gamma=\frac{|c_1|^2}{|c_0|^2+|c_1|^2} \frac{\omega^2}{\Omega} 
\label{Gamma_d'}
\ee
which is equal to (\ref{Gamma_d}) in the limit $\mu \ll \Omega$. Therefore both descriptions are compatible in this limit.
This result can be straightforwardly generalized to any finite number of sites coupled to a chain.

The model given by (\ref{c0_2_2}),(\ref{c1_2_2}) with $\gamma=0$ is a well-studied system whose eigenvectors and corresponding stability properties can be 
obtained analytically \cite{Ragh99}. In the particularly simple symmetric case with $\epsilon_0=0=\epsilon_1$ the eigenvalues and eigenvectors
are given by 
\be 
\mu=-J\eta/2+U(N-1)n\eta^2/(1+\eta^2)
\ee
with
\be
 \eta \in \{\pm 1\}, \quad |U(N-1)n| \le|J| 
\ee
and 
\be
\fl \qquad \eta \in \{\pm 1, -U(N-1)n/J \pm \sqrt{U^2(N-1)^2n^2/J^2-1} \}, \quad |U(N-1)n| \ge|J|,
\ee
where $n=|c_0|^2+|c_1|^2$ is the norm of the eigenvector and $\eta=c_0/c_1$ is real and nonzero. For a repulsive
interaction $U>0$ the anti-symmetric solution with $\eta=-1$ becomes dynamically unstable for $|U(N-1)n| \ge |J|$. 
The two solutions that only exist for sufficiently interactions are strongly localized in either the left or the right site, thus breaking
the symmetry of the system. 
\begin{figure}[htb]
\includegraphics[width=0.49\textwidth] {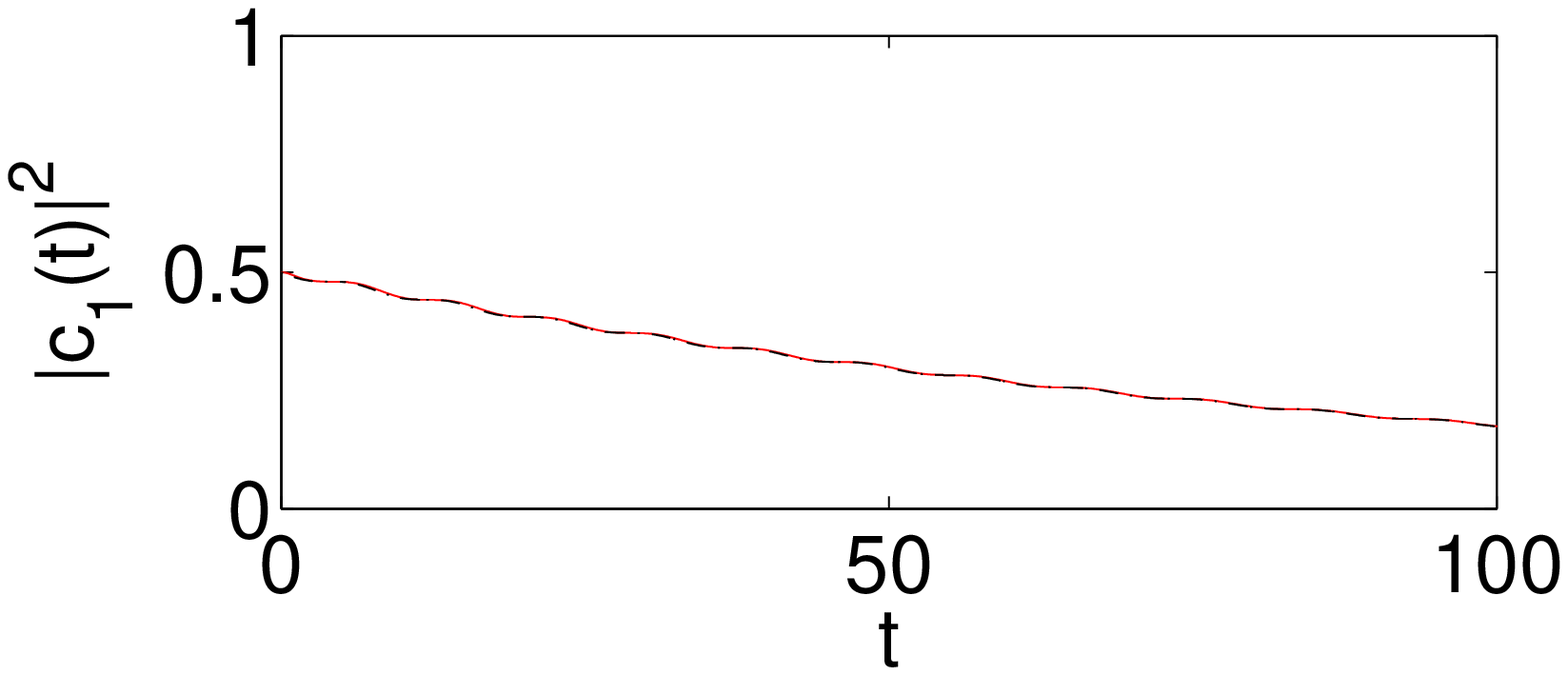}
\includegraphics[width=0.49\textwidth] {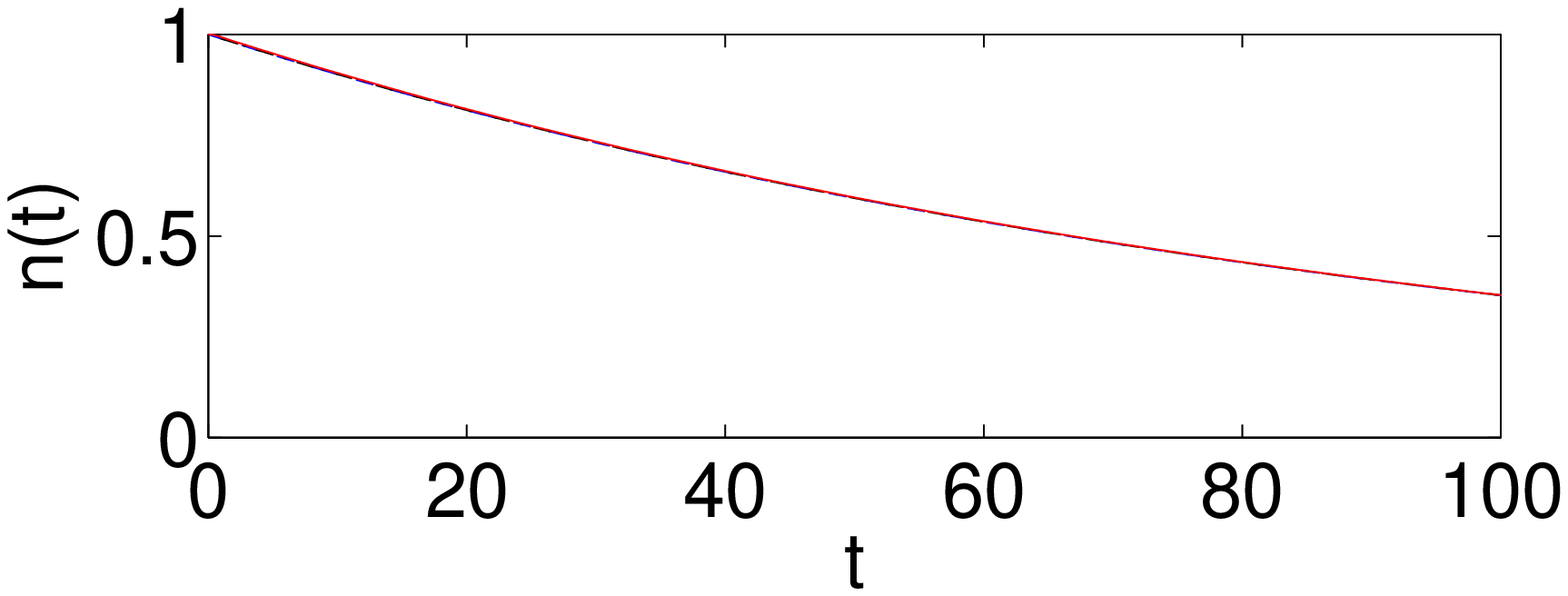}\\
\includegraphics[width=0.49\textwidth] {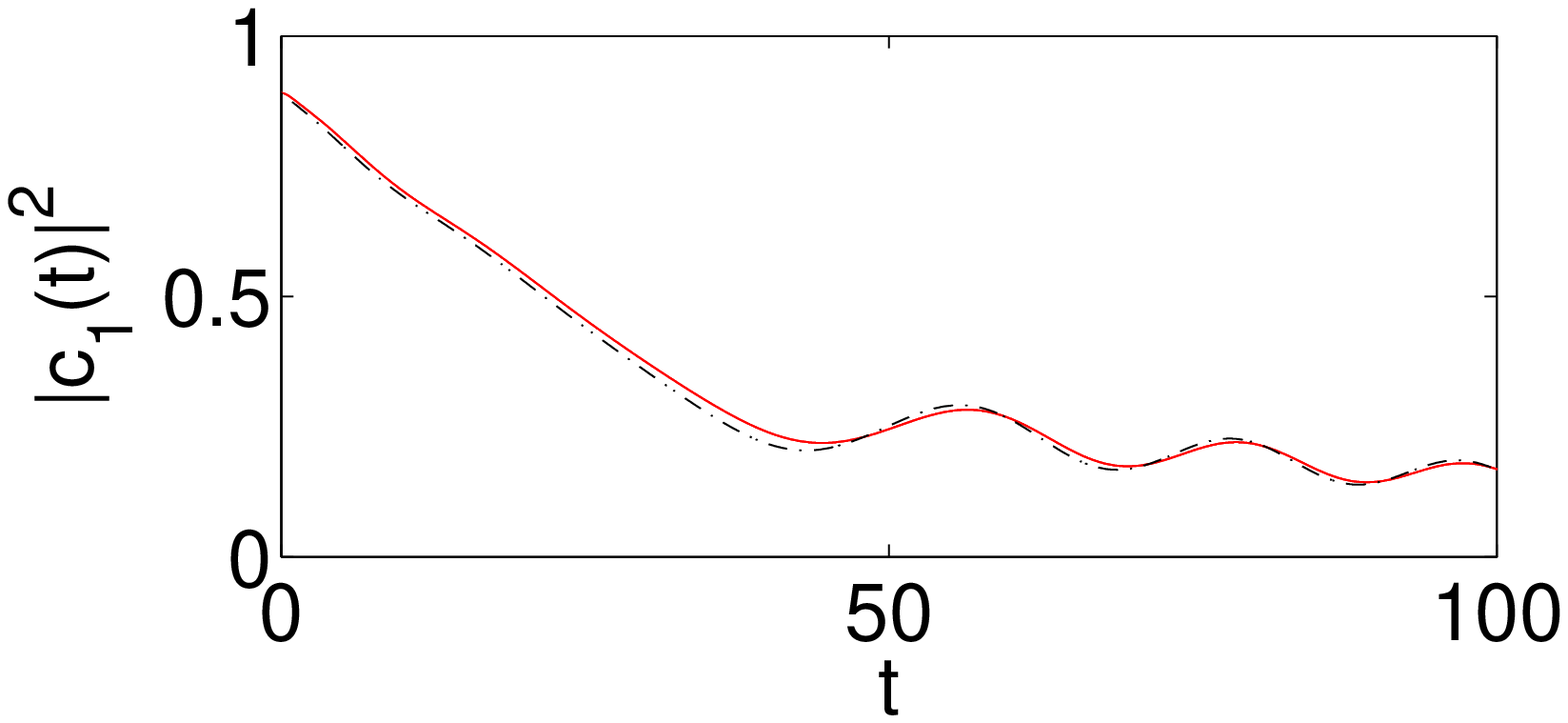}
\includegraphics[width=0.49\textwidth] {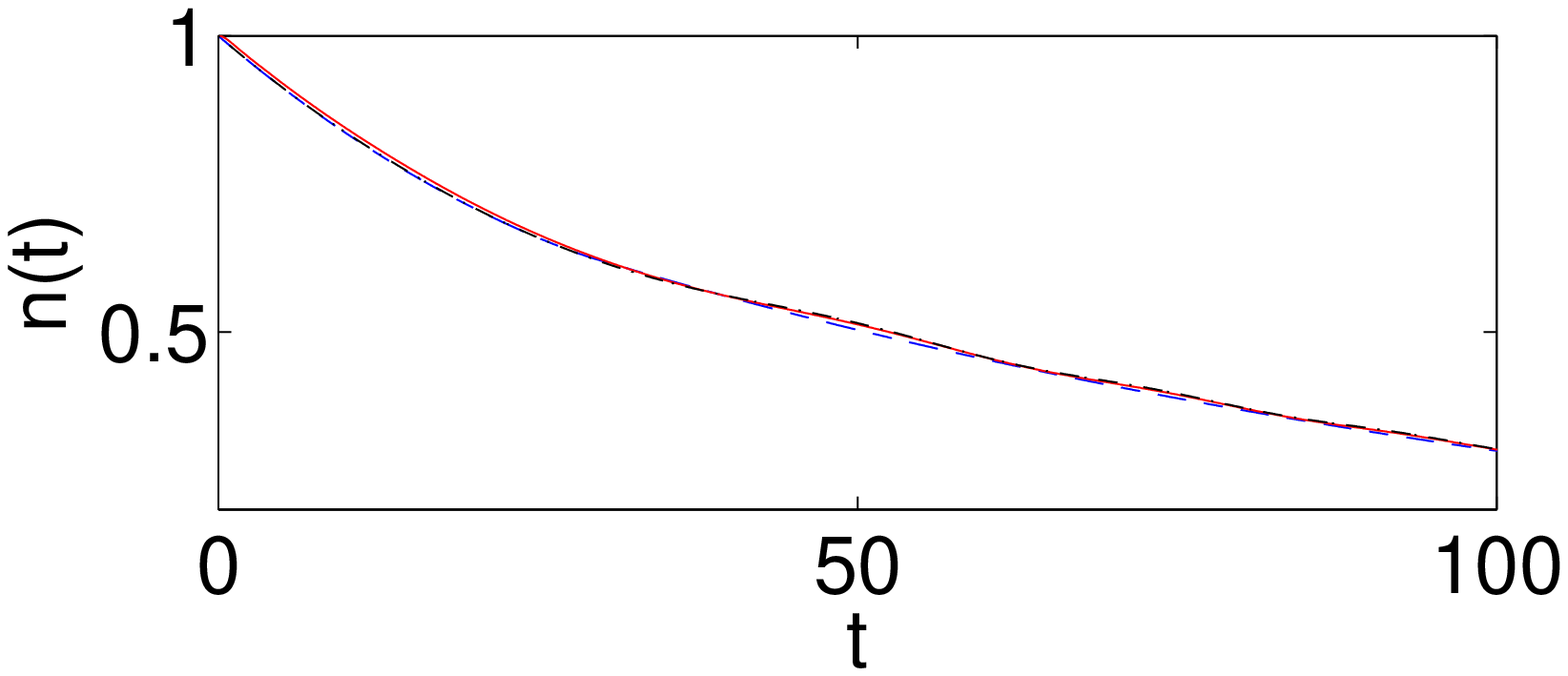}
\caption{\label{fig:double_stat} (Color online) Decay from a double well with parameters $\epsilon_1=0\epsilon_2$, $J=0.5$, $\omega=0.25$, $\Omega=3$, $\hbar=1$. $U(N-1)=0.8$
for the symmetric ground state (upper panels) and the eigenstate localized in the right well (lower panels).
The occupation $|c_1(t)|^2$ of the second lattice site and the occupation $n(t)=|c_0(t)|^2|+|c_1(t)|^2$ of the double well are shown as functions of time.
Predictions of the non-Hermitian two mode model (dashed dotted lines) are compared with a numerically exact propagation of the full nonlinear Schr\"odinger 
equation in a lattice with $200$ sites (solid lines). The dashed lines in the right panels correspond to quasistationary decay with decay rate $\Gamma_d'$ (\ref{Gamma_d'}).
}
\end{figure}
Figure \ref{fig:double_stat} shows the time propagation of two of the eigenstates of the symmetric double well due to the full 
mean-field dynamics of the whole system including the chain (solid lines) and according to the model (\ref{c0_2_2}), (\ref{c1_2_2})
(dashed dotted lines). The two approaches coincide well for both the symmetric ground state (upper panels) and the eigenstate 
localized in the right well (lower panels). For the symmetric ground we find an almost quasistationary decay behaviour since both the occupation 
$|c_1(t)|^2$ of the second lattice site and the total occupation $n(t)=|c_0(t)|^2|+|c_1(t)|^2$ of the double well show almost pure decay 
with only a small oscillation. Consequently the total occupation $n(t)$ is well described by a decay with rate $\Gamma_d'$ (dashed lines).
For the symmetry breaking state we observe a quasistationary decay until the effective interaction 
$U(N-1)n(t)$ drops below the threshold value $|J|$ for the existence of symmetry breaking eigenstates and the site 
occupations start to oscillate. In spite of this fact the total occupation $n(t)$ is still reasonably well described by a decay 
with rate $\Gamma_d'$ (dashed line). This behaviour is in agreement with previous studies of open double well systems containing 
detailed discussions of the dynamics of the (quasi-) eigenstates \cite{Witt08,Trim11,08nhbh_s,09ddshell}. 
\begin{figure}[htb]
\includegraphics[width=0.49\textwidth] {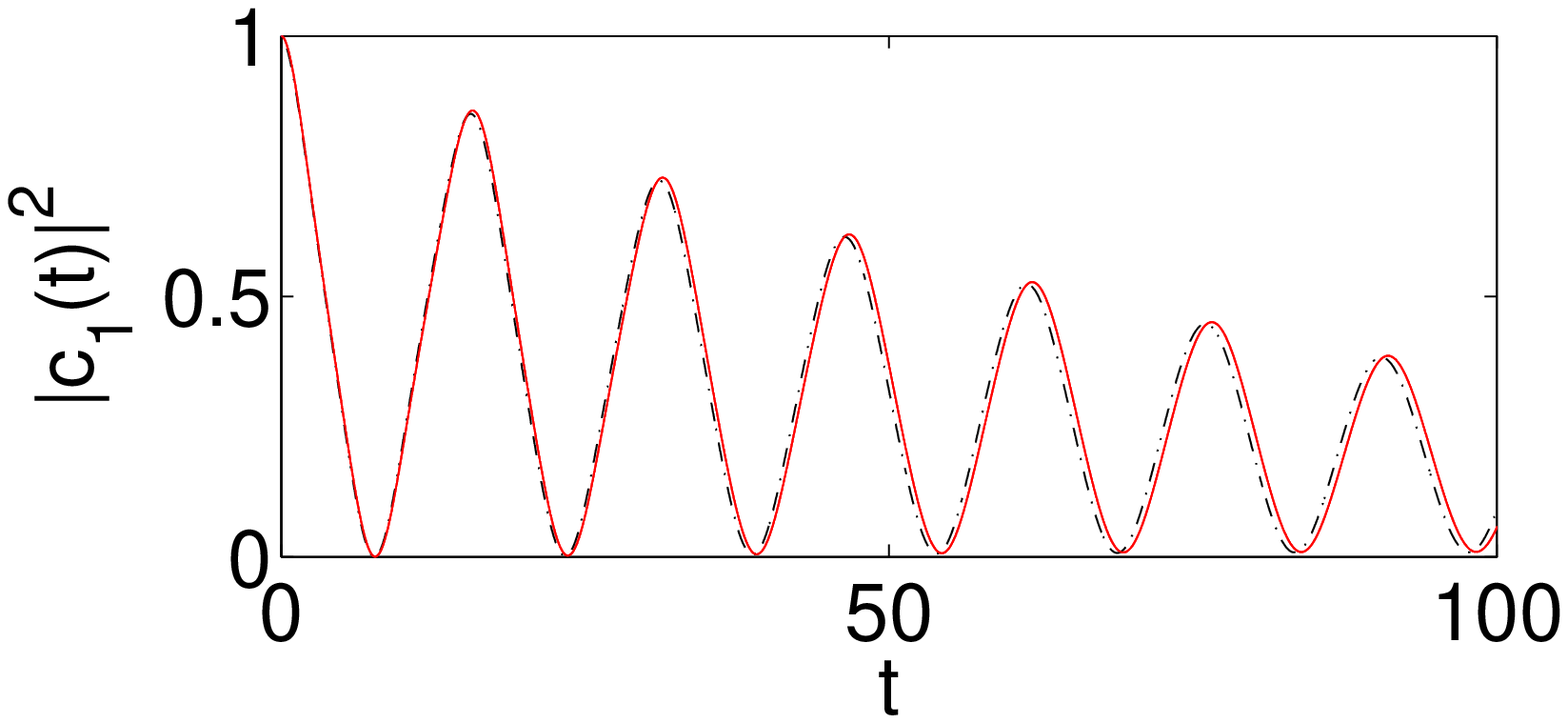}
\includegraphics[width=0.49\textwidth] {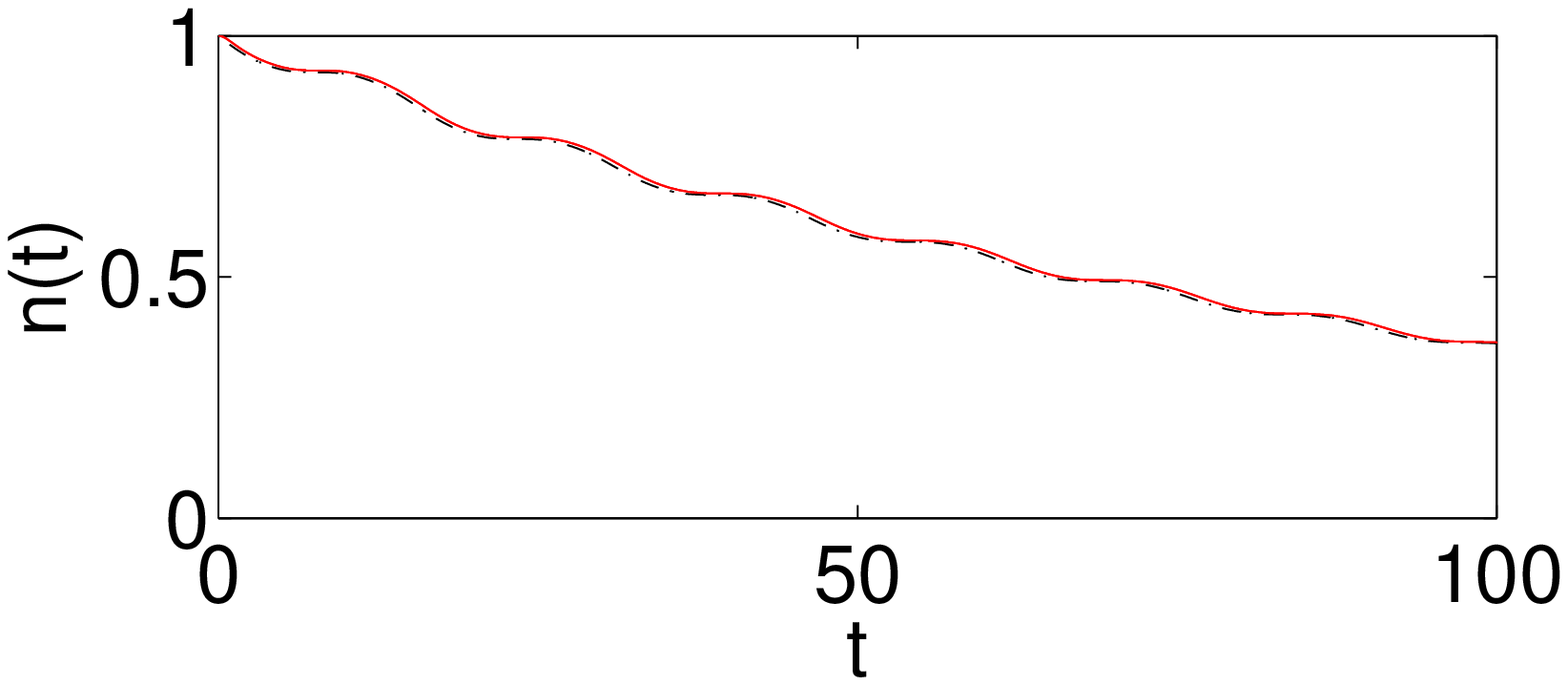}\\
\includegraphics[width=0.49\textwidth] {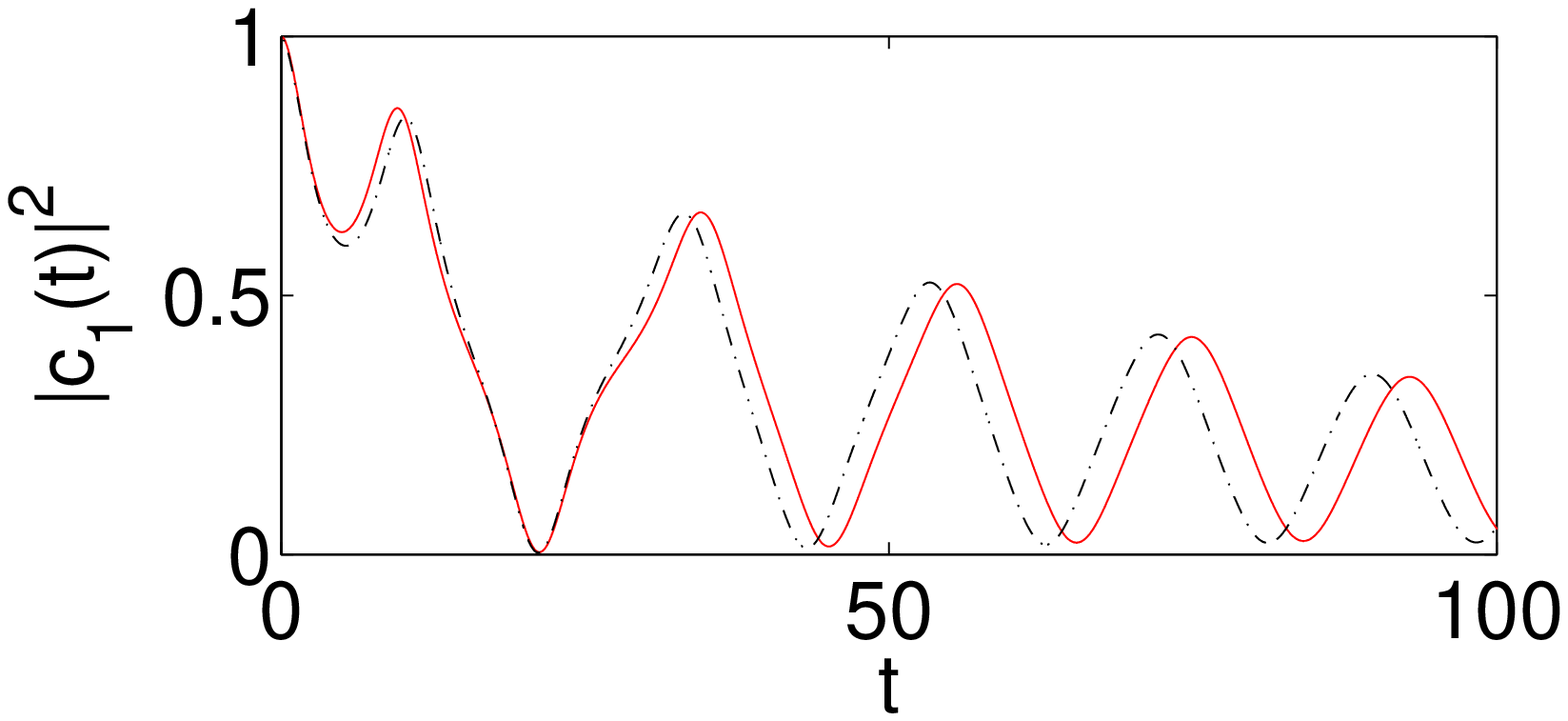}
\includegraphics[width=0.49\textwidth] {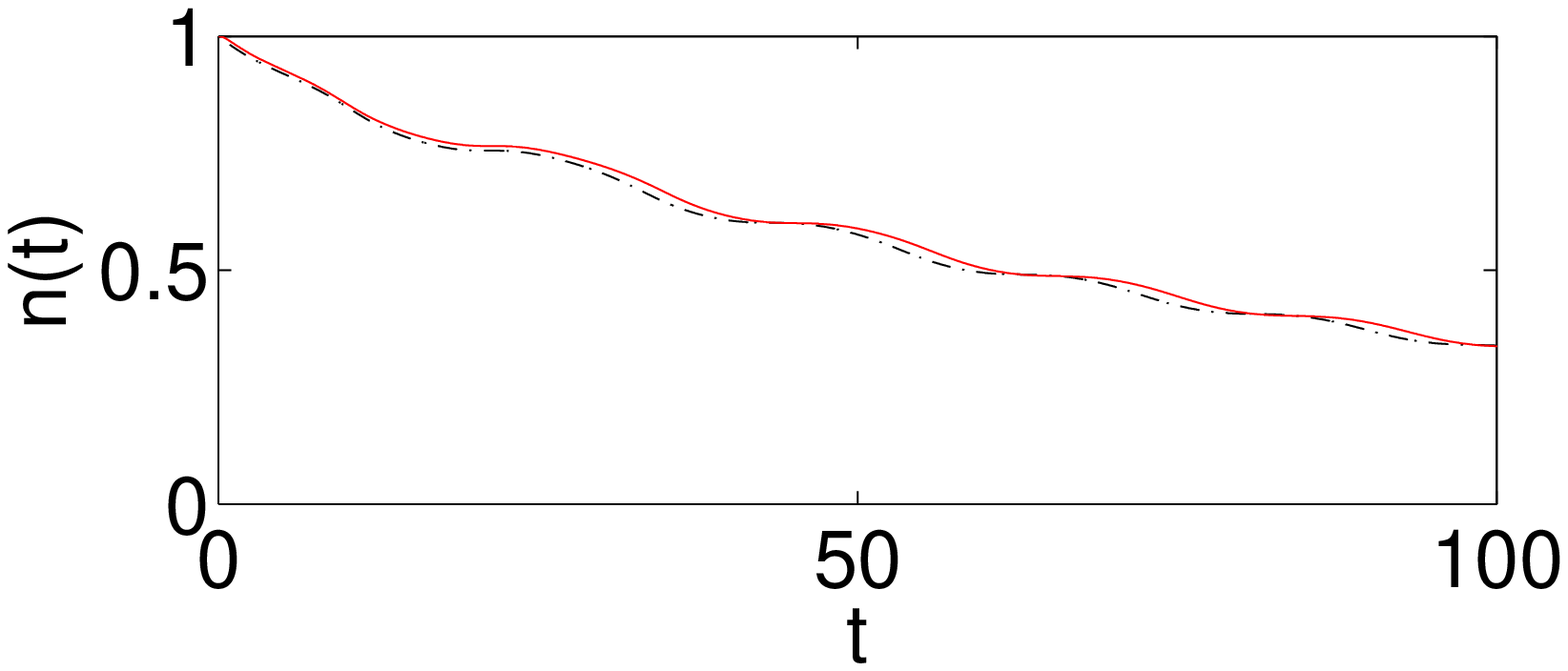}
\caption{\label{fig:double_dyn} (Color online) Decay from a double well with parameters $\epsilon_1=0\epsilon_2$, $J=0.5$, $\omega=0.25$, $\Omega=3$, $\hbar=1$, $U(N-1)=0.8$ (upper panels),
$U(N-1)=1.1$ (lower panels) for the initial conditions $c_0(0)=0$, $c_1(0)=1$. The relative occupation $|c_1(t)|^2$ of the second lattice site and the relative occupation 
$n(t)=|c_0(t)|^2|+|c_1(t)|^2$ of the double well are shown as functions of time.
Predictions of the non-Hermitian two mode model (dashed dotted lines) are compared with a numerically exact propagation of the full nonlinear Schr\"odinger 
equation in a lattice with $200$ sites (solid lines).
}
\end{figure}
Figure \ref{fig:double_dyn} shows the time propagation for the initial conditions $c_0(0)=0$, $c_1(0)=1$, i.e.~all particles 
are initially in the right well. For a moderate value $U(N-1)=0.8$ of the interaction (upper panels) the oscillatory behaviour of the full system 
(solid lines) is well capturd by the non-Hermitian two mode model (\ref{c0_2_2}), (\ref{c1_2_2}) (dashed dotted lines). 
The lower panels display the dynamics for a higher value $U(N-1)=1.1$ of the interaction which lies above the treshold 
$U(N-1)=2|J|=1$ for running phase self-trapping (see e.g.~\cite{Ragh99,Kalo03,Kalo03a,Mors06,Fu06,12heur}) where interactions prevent a total population transfer between 
the wells in the corresponding hermitian double well system. Such an oscillation with a small amplitude can also be observed here for 
short times. However, after the first oscillation the effective interaction $U(N-1)n(t)$ drops below the threshold value $2|J|=1$ 
so that oscillations with larger amplitudes are possible again. Before the threshold is reached, the non-Hermitian two mode model (dashed dotted lines) provides an excellent description of
the dynamics of the full system (solid lines), afterwards deviations occur but the qualitative behaviour ist still correctly described.
 
Now we again turn to the dynamics of the full Bose-Hubbard system and compare it with an effective two mode description of our open 
double well, namely the Lindblad master equation \cite{Breu02}
\be
   \dot{{\hat \rho}}=-\ri[{\hat H}_0,{\hat \rho}]-\frac{1}{2}\gamma \left( {\hat a}_1^\dagger {\hat a}_1{\hat \rho}+ {\hat \rho}{\hat a}_1^\dagger {\hat a}_1-2 {\hat a}_1 {\hat \rho}{\hat a}_1^\dagger \right)
\label{Lindblad}
\ee
for the density matrix $\rho$ where the second term describes constant decay from site $1$ with rate $\gamma$ whereas the first term provides the hermitian part  
of the time evolution with the two site Bose-Hubbard Hamiltonian 
\be 
 {\hat H}_0= -(J/2) ({\hat a}_1^\dagger {\hat a}_0+{\hat a}_0^\dagger {\hat a}_1)+ (U/2) ({\hat a}_0^{\dagger2}{\hat a}_0^2+{\hat a}_1^{\dagger2}{\hat a}_1^2) \,.
\ee
The time-dependent expectation values of an operator ${\hat O}$ is then given by the trace 
$\langle {\hat O}(t) \rangle={\rm Tr(\rho(t){\hat O})}$. Recently it was shown \cite{Witt08,Trim11} that the Lindblad master equation (\ref{Lindblad}) reduces to the nonlinear non-Hermitian model 
(\ref{c0_2_2}), (\ref{c1_2_2}) in the mean-field limit. 
\begin{figure}[htb]
\includegraphics[width=0.49\textwidth] {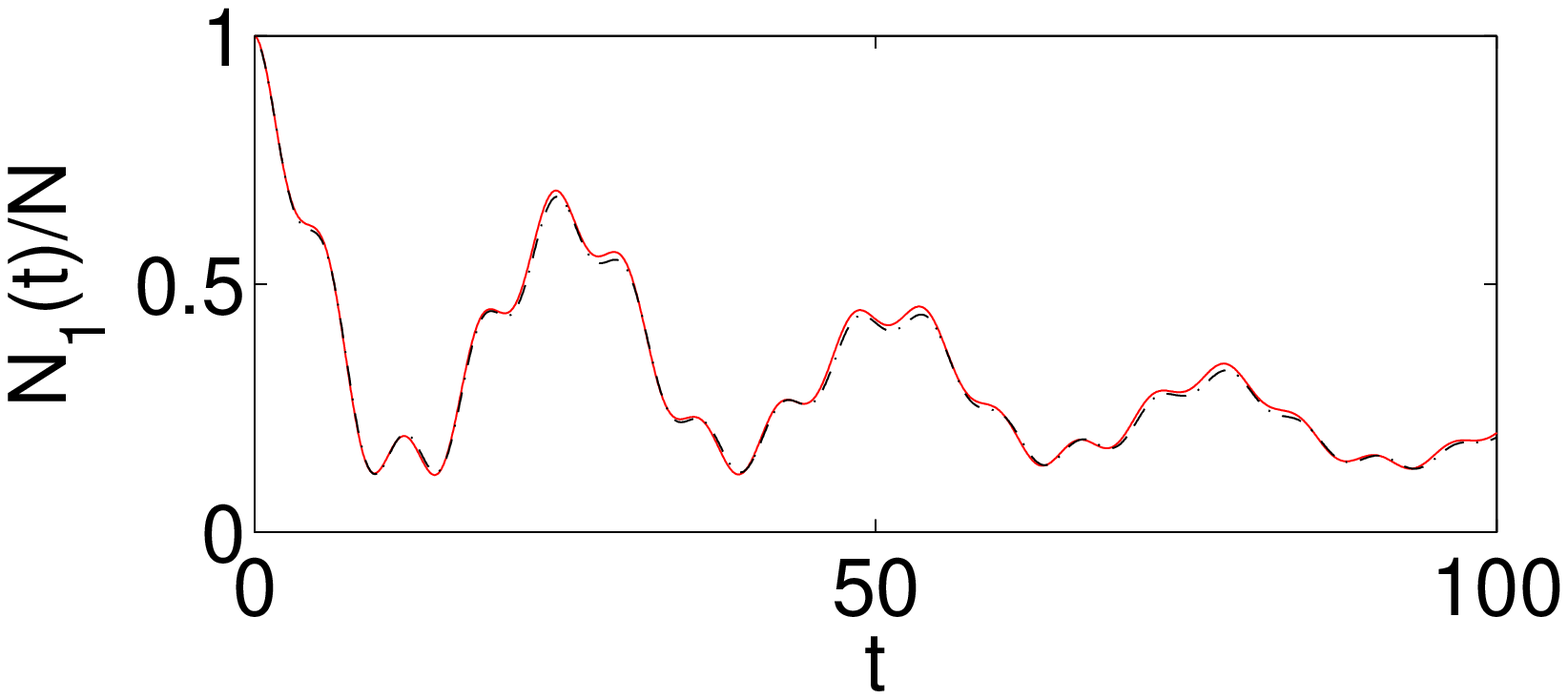}
\includegraphics[width=0.49\textwidth] {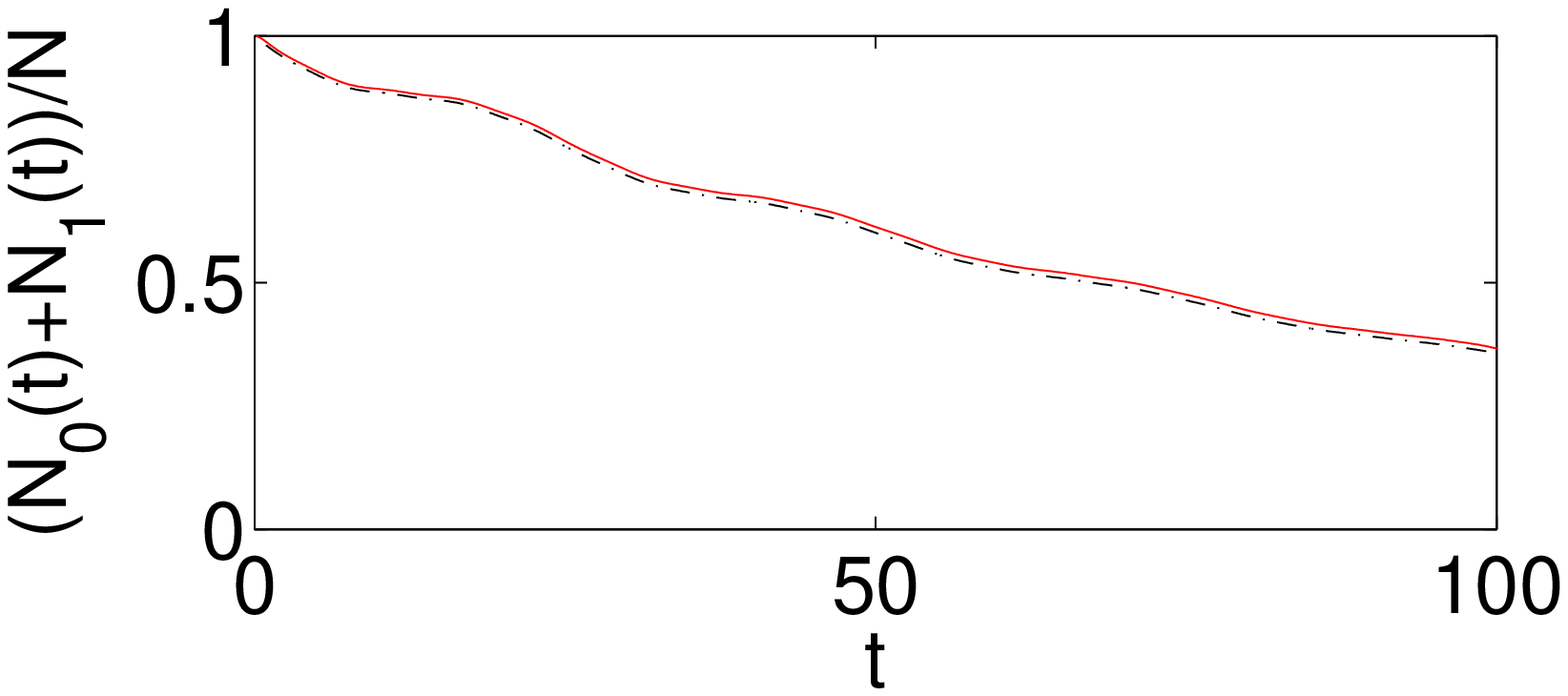}\\
\includegraphics[width=0.49\textwidth] {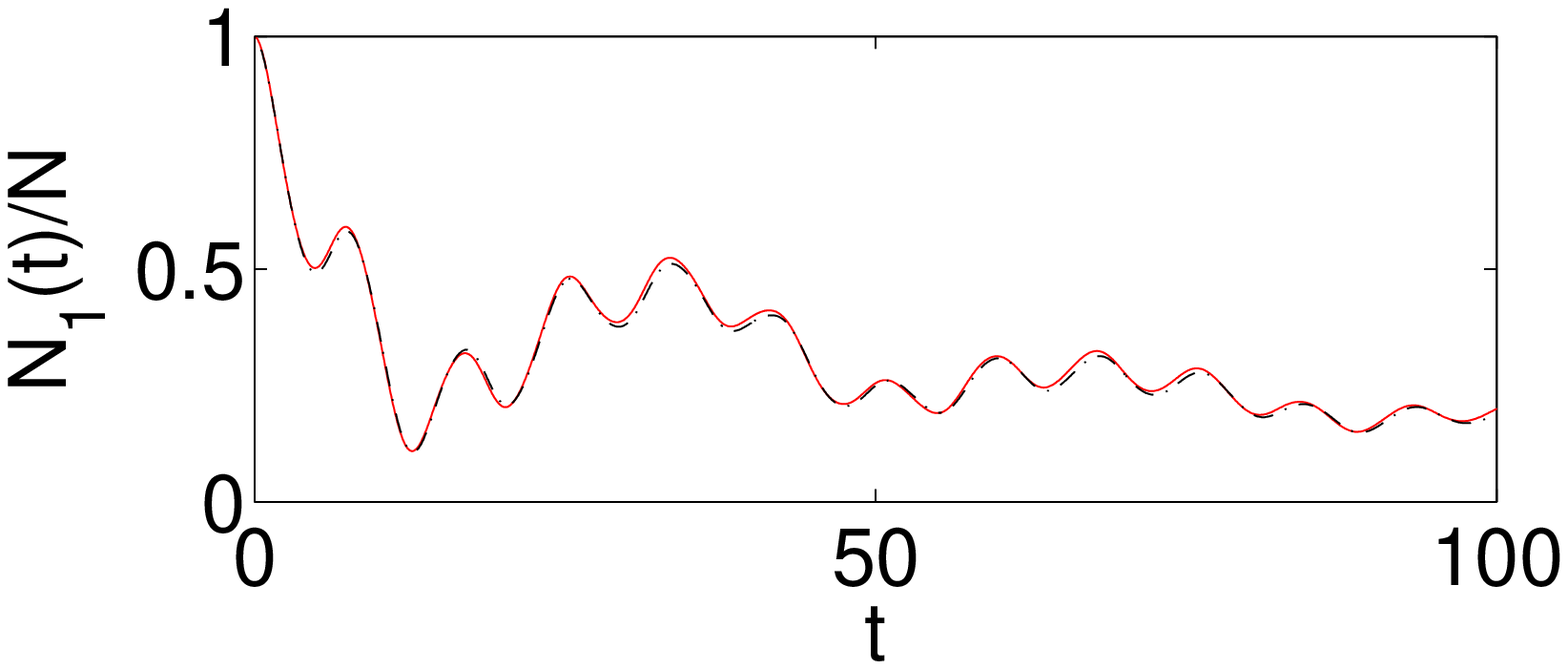}
\includegraphics[width=0.49\textwidth] {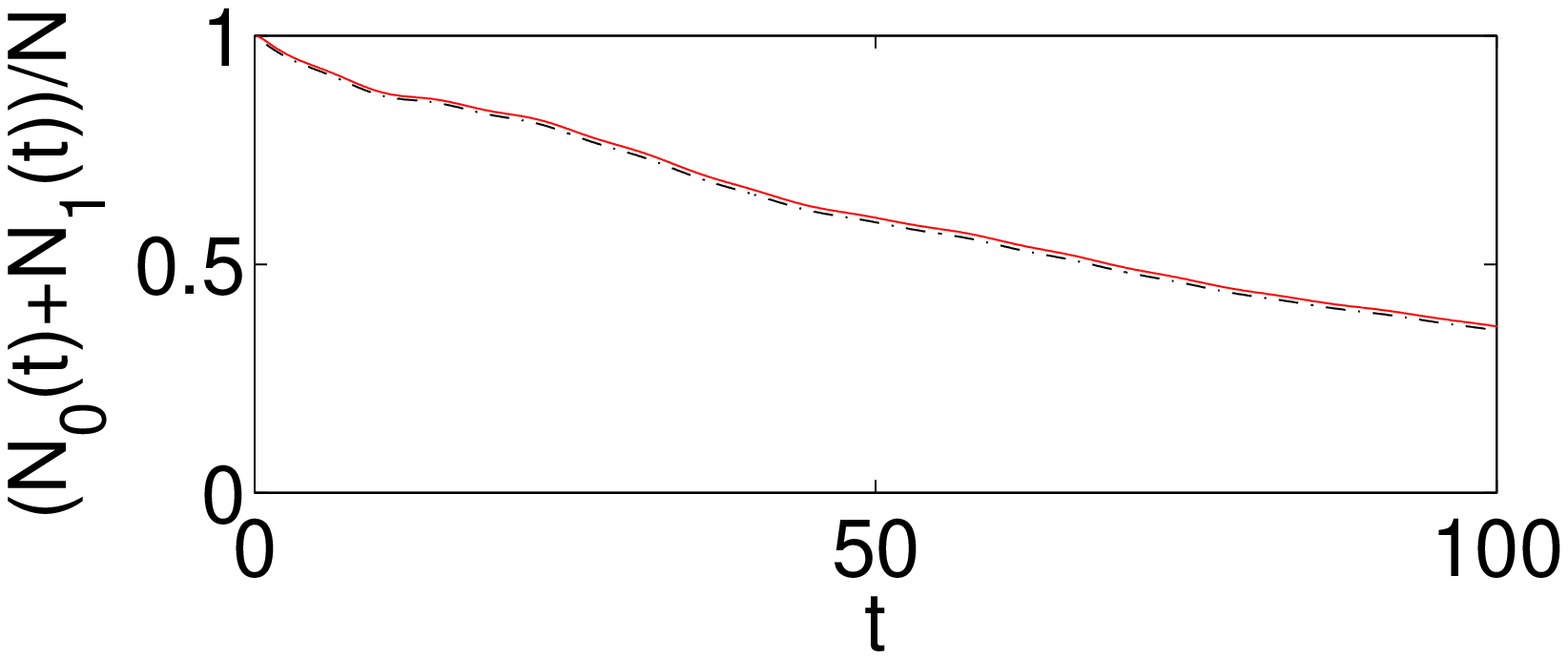}
\caption{\label{fig:double_MP} (Color online) Decay from a single site with parameters $\epsilon_1=0\epsilon_2$, $\omega=0.25$, $\Omega=3$, $\hbar=1$,$U(N-1)=0.8$
for initial conditions where only the second well is occupied at $t=0$. The relative occupation $N_1(t)/N$ of the second lattice site and the relative occupation $(N_0(t)+N_1(t))/N$ of the double well are shown as functions of time. 
Predictions of the Lindblad master equation (\ref{Lindblad}) (dashed dotted lines) are compared with a numerically exact propagation of the full Bose-Hubbard Hamiltonian in a lattice with $150$ sites (solid lines).
Upper panels: $N=2$ particles, lower panels: $N=3$ particles.
}
\end{figure}
For a small number of particles, equation (\ref{Lindblad}) can be integrated directly, for higher particle numbers not considered
here it can be solved using Monte Carlo methods \cite{Breu02}.  
Figure \ref{fig:double_MP} displays the relative occupation of the second lattice site $N_1(t)/N$, $N_1(t)={\rm Tr(\rho(t){\hat a}_1^\dagger {\hat a}_1 )}$ (left panels) 
and the relative total occupation $(N_0(t)+N_1(t))/N$ of the double well (right panels) for initial conditions where only the second well is occupied at $t=0$. 
Both for $N=2$ particles (upper panels) and for $N=3$ particles the full Bose-Hubbard dynamics (solid lines) is well reproduced
by the Lindblad master equation (dashed dotted lines). Compared to the mean-field case the oscillatory dynamics of the system appears
less regular due to the occurrence of various frequencies corresponding to transitions between different eigenstates
of the many- (or here rather few-) particle system, an effect which, if regarded from a mean-field point of view, is also referred to as 
quantum fluctuations in this context \cite{Kalo03a,Kalo03,12heur}.

\section{Conclusion}
\label{sec:conclusion}
In this article, an experimentally realizable system, namely cold bosons in an optical lattice modeled by a Bose-Hubbard Hamiltonian 
was used to theoretically investigate the dynamics of open interacting many-particle systems within a closed system setup.
Open single and double well systems were simulated as one respectivelly two sites weakly coupled to a long but finite Bose-Hubbard 
chain avoiding additional approximations due to absorbing boundaries, complex absorbing potentials or the introduction of baths.

Even for single site tunnelling deviations of the mean-field dynamics from the full Bose-Hubbard dynamics were found in accordance 
with reference \cite{Lode09}. The non-exponential decay behaviour and in particular the differences between Bose-Hubbard
and mean-field were demonstrated to depend on the dispersion relation in the chain. Both in the mean-field and many-particle case the 
dynamics was well described by rate equation models derived using a discrete lattice version of the Siegert approximation method.

It was shown that a description of tunnelling decay by means of constant local decay terms
is justified if the chemical potential of the considered subsystem is small compared to the tunnelling coefficient in the lattice. 
In the latter limit the dynamics of an open double well was analyzed. 
For the latter it was found that full mean-field and Bose-Hubbard dynamics is well described by a non-hermitian nonlinear 
Schr\"odinger equation respectivelly a Lindblad master equation with a constant decay term.

In summary, decaying interacting quantum systems were analyzed for a concrete physical situation by means of a closed system approach, explicitly
confirming the validity of popular effective theoretical descriptions that are usually applied on a phenomenological basis.

\section*{Acknowledgment}
The author would like to thank H. J. Korsch, E. M. Graefe and M. P. Strzys for helpful discussions and comments.
  
\section*{References}

\end{document}